# Design of a full-filed transmission X-ray microscope with 30nm resolution

Keliang Liao[1, a *], Qili He[2], Panyun Li[1], Maohua Song[1], Peiping Zhu[1,2,b*]

[1]Jinan Hanjiang Opto-electronics Technology Ltd., Shandong, 250000, China

[2]Institute of High Energy Physics, Chinese Academy of Sciences, Beijing 100049, China

[a]hjoptech@163.com, [b]zhupp@ihep.ac.cn

**Abstract.** A full-field transmission hard X-ray microscope (TXM) with 30nm resolution was designed and its prototype was constructed. The TXM relies on a compact, high stiffness, low heat dissipation and low vibration design philosophy and utilizes Fresnel Zone plate (FZP) as imaging optics. The design of the TXM was introduced in detail, including the optical layout, the parameters of the FZP, the mechanical design of the TXM instrument. Preliminary imaging result with 52nm spatial resolution was achieved.

## 1. Introduction

The microscope is the eye for human observing the microcosm. Throughout the long period of development from the light microscope to the electron microscope, each significant invention in microscopy has brought a profound impact on the advancement of science and technology. X-ray microscopy represents another milestone, serving as a new window for people to observe the microcosm and becoming a new focus for scientific development. Compared to visible light, X-rays have a wavelength three orders of magnitude smaller, allowing for a three-order-of-magnitude improvement in resolution limited by diffraction[1-3]. In comparison to electron beams, X-rays possess powerful penetrating capabilities, enabling the non-destructive acquisition of density distribution information with nanoscale spatial resolution inside samples.

Full-field transmission X-ray microscope (TXM) has the capability to observe the three-dimensional structure of samples at the nanoscale[4-7]. TXM presents enormous application prospects and has the potential to become as ubiquitous as optical and electron microscopes, finding its way into the laboratories of scientists and engineers, serving as an indispensable research tool in fields such as life sciences, materials science, energy science, information science, and environmental science. Due to the importance and significant application prospects of TXM, a lot of breakthroughs have been achieved over the past decade. Developed countries around the world have successively established X-ray TXM beamlines and experimental stations at synchrotron facilities, such as ESRF, APS, NSLS, and Spring8[3, 8]. Many X-ray TXM beamlines and experimental stations at synchrotron facilities have also been established in China[9-11].

In addition to synchrotron-based TXM, significant progress has also been made in TXM based on X-ray tubes. As one of the most advanced scientific instruments in the world, TXM involves multiple cutting-edge technologies, including high-brightness X-ray sources, high-resolution X-ray optical components, high-precision CT sample stage system, and high-precision environmental control system. The TXM industrial project conducted in Jinan Hanjiang Opto-electronics Technology Ltd. (HJ OPTECH) is driven by the needs of a diverse industrial community preferring a non-destructive high-resolution microscope, coinciding well with several key industries in Shandong Province. In this work, we introduced the design of a full-field TXM with 30nm resolution based on the micro-focus X-ray tube. First, we briefly discuss the optical layout, the corresponding optical design and the mechanical design of the TXM. Second, we reported the preliminary imaging result of the TXM instrument. The focusing performance of the condenser and the spatial resolution test were both discussed. We close with an outlook of the future work.

## 2. TXM Design

### 2.1. Optical layout

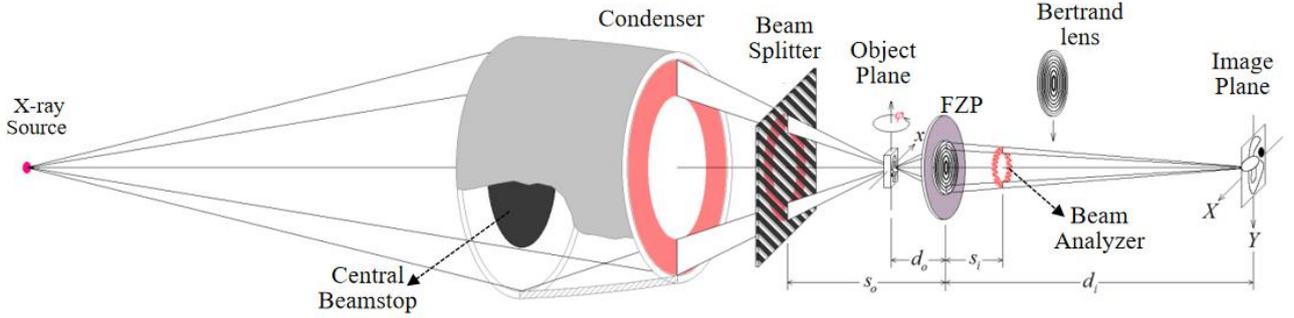

Fig.1. Optical layout of hard x-ray nano-CT system.

The basic principle of the TXM instrument is the full-field microscopic imaging by utilizing FZP as the magnification lens, as shown in Fig.1. The instrument is equipped with several components such as X-ray source, condenser, beam splitter, sample, FZP, beam analyzer, Bertrand lens and imaging detector.

The X-ray emitted from the X-ray tube is focused by the condenser. As presented in Fig.1, in the front center of the condenser, a beamstop (BS) is utilized to absorb the central X-rays, and provides a hollow cone illumination beam for the TXM. With this dedicated hollow cone beam design, the illumination beam and the image could be separated from each other in the imaging plane. The illumination beam passes through the beam splitter and irradiates on the sample. The modulated X-rays, which contain the information of the sample, is then magnified by the FZP. The detector will record the amplified projections modulated by the sample in the image plane. The beam analyzer is utilized together with the above beam splitter in the phase contrast imaging mode.

Unlike the X-ray nanoprobe, which could only obtain the sample information point by point, the TXM can acquire the whole image of the sample at one time. When combined with CT technology, the TXM, equipped with FZPs with different outermost ring width and different diameter, can realize multi-scale non-destructive three-dimensional imaging of samples with a diameter of 5μm-100μm and a resolution of 30nm-150nm.

In theory, with the modulation of the sample, totally five major imaging contrast could be provided in the X-ray imaging regime, including the traditional absorption image, the phase image, the differential phase contrast image, the phase Laplacian image, and the scattering image. This TXM instrument is majorly developed for the absorption imaging, the differential phase contrast imaging and the scattering imaging.

### 2.2. Parameters of FZP

The imaging schematic of the FZP was shown in Fig.2(a). The FZP could be viewed as a lens, which should work according to the basic principle of the geometrical optics, with the distance between the object and the FZP being denoted as $d_0$, and the distance between the FZP and the image being denoted as $d_i$. The FZP is a diffraction optical element, with a series of zones satisfying the zone plate law, as depicted in Fig.2(b).

An X-ray tube with the chromium (Cr) target was utilized. The Cr target has a characteristic peak energy of 5.41 keV, with Kα1 characteristic peak energy at 5.414 keV and Kα2 characteristic peak energy at 5.405 keV. In order to diminish the chromatic aberration, the zone number (N) of FZP should satisfy the following relationship,

$$N = \frac{E}{\Delta E} = \frac{5.41}{5.414 - 5.405} = \frac{5.41}{0.009} = 601. \qquad (1)$$

Based on the chromatic aberration limit, the FZP with a focal length of 8.7mm and an outer ring width of 28.5nm has been chosen. The calculated zone number is slightly greater than the above limit, as shown in the following equation,

$$N = \frac{F\lambda}{4\Delta r_N^2} = \frac{8.7mm \times 0.23nm}{4 \times (28.5nm)^2} = \frac{8.7 \times 10^6 \times 0.23}{4 \times (28.5)^2} = 616. \quad (2)$$

Then the corresponding diameter of the FZP is derived,

$$D = 2\sqrt{N\lambda F} = 2 \times \sqrt{616 \times 0.23nm \times 8.7mm} = 70\mu m. \quad (3)$$

For the TXM instrument, the gold-structure zone plate with an outer ring width of 28.5nm was designed. Due to manufacturing process limitations, the maximum aspect ratio of the gold grating bars on the zone plate is approximately in the range of 20 to 30, with a designed thickness of 800nm. The parameters of the FZP are listed in Table 1, and a SEM photo of the gold FZP is shown in Fig.2(c).

Table 1. The parameters of this zone plate

| Energy (keV) | Material | Diameter (μm) | Outer ring width (nm) | Number of zones | Height (nm) | Focal length (mm) |
|---|---|---|---|---|---|---|
| 5.41 | Au | 70 | 28.5 | 616 | ≥800 | 8.7 |

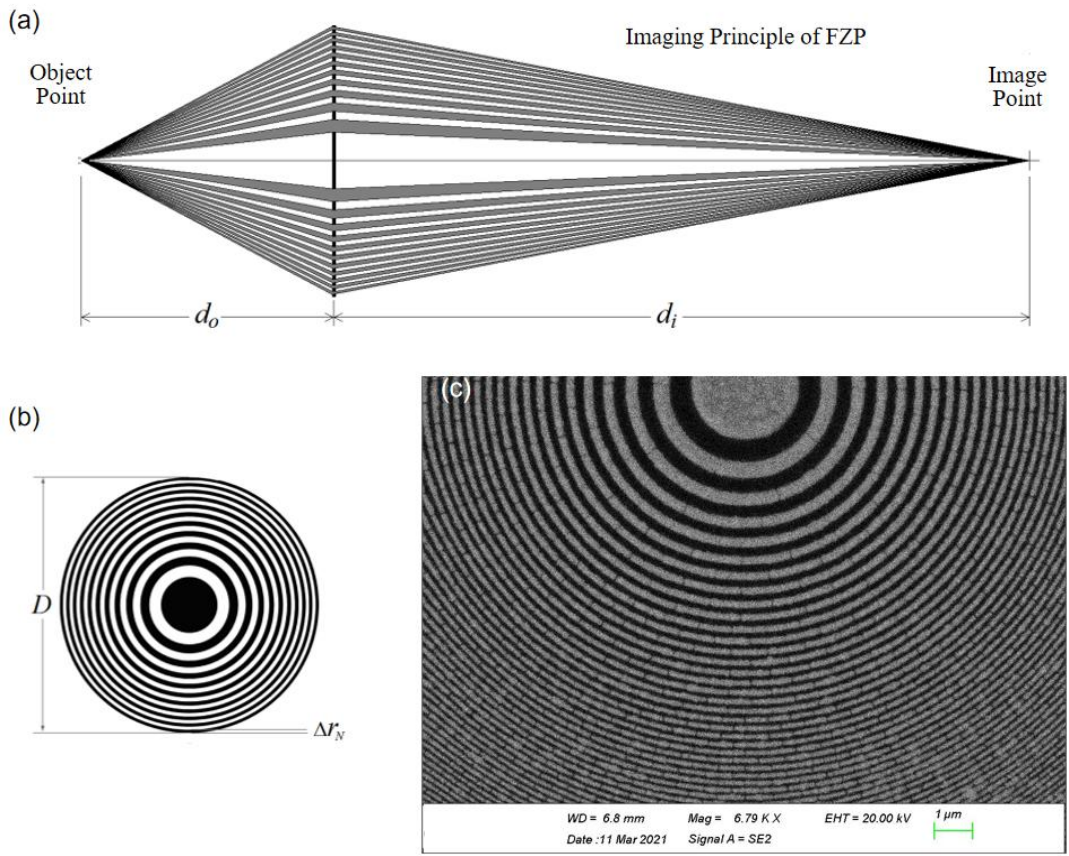

Fig2. The principle of Fresnel zone plate as an imaging lens.

## 2.3 Imaging Parameters

For the TXM instrument, the imaging resolution $\hat{\delta}$ is 30nm, the depth of field (DOF) of the zone plate is

$$DOF = \frac{3.28}{\lambda}\hat{\delta}^2 = 12.835\mu m, \quad (4)$$

The detector with 7.52μm pixels size and 2084 pixels is used in this project. An optical coupling system with about totally 15x magnification is placed between the scintillator screen and the detector sensor. The magnification of FZP is chosen to be 50x, resulting in a total magnification of

750x for the TXM. Under this configuration, the pixel size of the detector sensor is 7.52μm, which corresponds to a pixel size of 500nm on the scintillator screen. On the object plane, the corresponding pixel size is 10nm. The field of view captured on the object plane is 20.8μm. Table 2 lists the relative magnifications for the TXM instrument.

Table 2. The relative magnifications for the TXM instrument

| Pixel size on the object surface | Magnification factor of zone plate | Pixel size on the scintillator screen | Magnification factor of optical system | Pixel size of the CCD sensor | Total magnification factor |
|---|---|---|---|---|---|
| 10nm | 50 | 500nm | ~15 | 7.52μm | 750 |

Referring to Fig.2, assuming M is the magnification factor of the FZP, the object distance can be determined as follows:

$$d_o = \frac{M+1}{M}F = \frac{51}{50} \times 8.7\text{mm} = 8.874\text{mm}, \quad (5)$$

And the imaging distance is:

$$d_i = (M+1)F = 51 \times 8.7\text{mm} = 443.7\text{mm}. \quad (6)$$

## 2.4 Mechanical Design

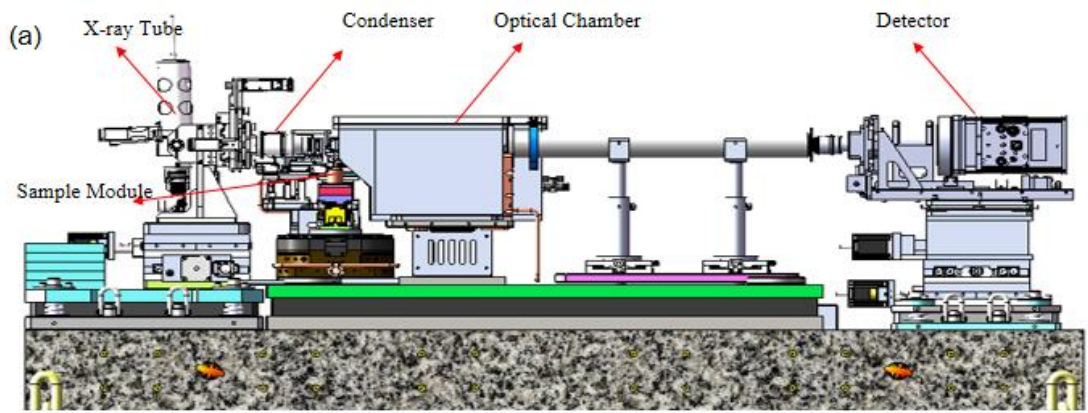

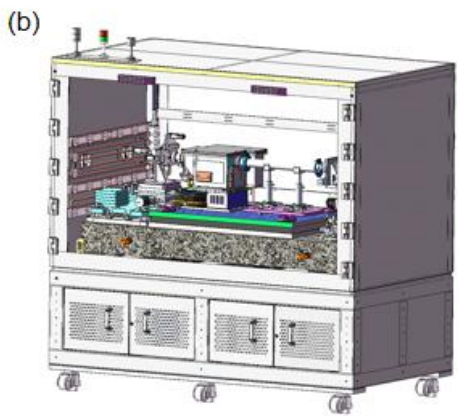

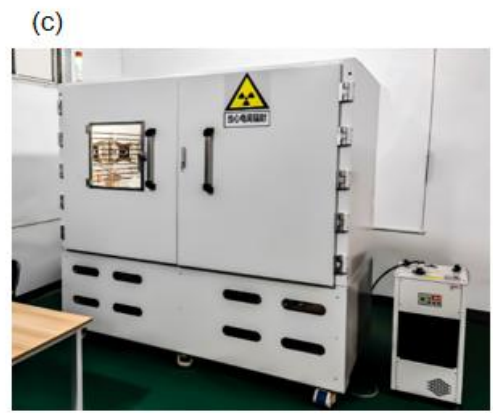

Fig 3. The schematic and photograph of the TXM. (a) The 3D model of the major modules. (b) The 3D schematic of the TXM. (c) The photograph of the TXM instrument.

The prototype for the TXM as shown in Fig.3 has been constructed and tested. For the first phase of the TXM industry project, totally two sets of prototypes were fulfilled. The TXM consisted of several important modules, including the X-ray tube, the condenser module, the sample module, the optical chamber, and the detector module, which were clearly marked in Fig.3(a). All the above major modules of the TXM were mounted on a dedicated marble supporting base. Moreover, the major modules of the TXM, together with the marble supporting base were mounted on a welding supporting frame with high rigidity, as shown in Fig.3(b). This TXM instrument was totally more than 4 tons, and its photograph was depicted in Fig.3(c).

## 3. Preliminary imaging result

Preliminary imaging results with about 52nm spatial resolution were achieved, including the focusing performance of the condenser, the imaging test of tungsten nanoparticles.

First, as shown in Fig.4(a), a scintillator plate was placed near the focus of the condenser, which could transfer the X-ray beam to visible lights. A simple imaging system with a Navitar Zoom6000 module and a Hamamasu CMOS detector were utilized to observe the 2D intensity distribution of the condenser's focus. By scanning the above imaging system along the optical axis, the focus point was finally determined. The intensity distribution of the focus was depicted in Fig.4(b). Moreover, Fig.4(c) showed the 1D shape of the focus in the horizontal direction, which was fitted with the Gaussian function, and a focus size of 26.1μm FWHM was calculated.

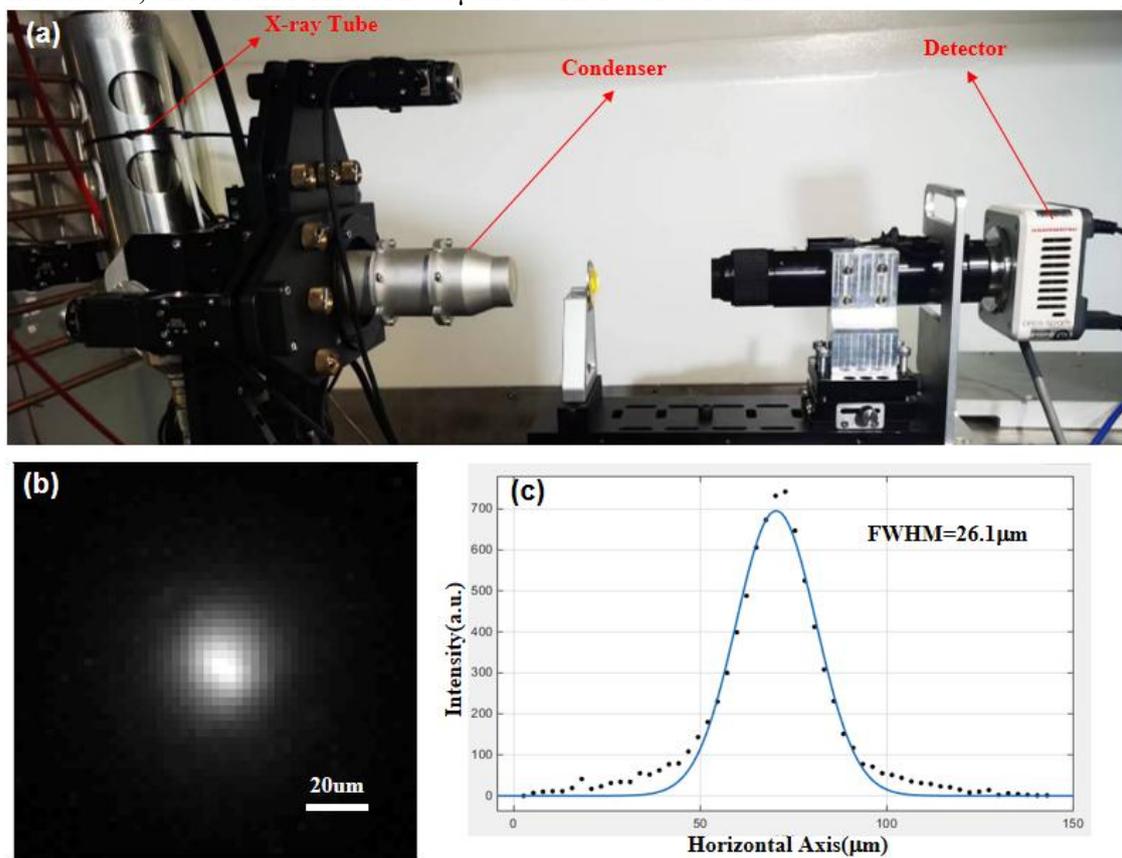

Fig 4. Characterization of the illumination beam focused by the condenser. (a) The photograph of the experiment setup. (b) The 2D intensity distribution in the focal plane. (c) The 1D shape of the focus in the horizontal direction.

Second, the optical chamber with FZP mounted inside and the optical coupling X-ray detector with high resolution were carefully calibrated into the optical path, as shown in Fig.5(a). For a demonstration imaging experiment, a few tungsten nanoparticles with diameter about 50nm were randomly sprayed onto a Kapton membrane with 10μm thickness. Furthermore, the Kapton membrane with nanoparticles was transferred to the sample holder, as shown in Fig.5(b). A typical random distribution of tungsten nanoparticles was observed by using SEM, which was provided by the supplier and depicted in Fig.5 (c). After a series of adjustment of the sample and the FZP, a clear image of the nanoparticles in absorption mode was acquired, as shown in Fig.5 (d). A lot of nanoparticles could be observed, among which several nanoparticles may join together and form irregular nanoclusters, as magnified in Fig.5(e). A typical nanoparticles was marked with red line L1 and its transmission curve along the horizontal direction was depicted in Fig.5(f), which coincided well with a nanoparticle with about 50nm diameter. Moreover, another red line L2 was also selected and presented in Fig.5(g).

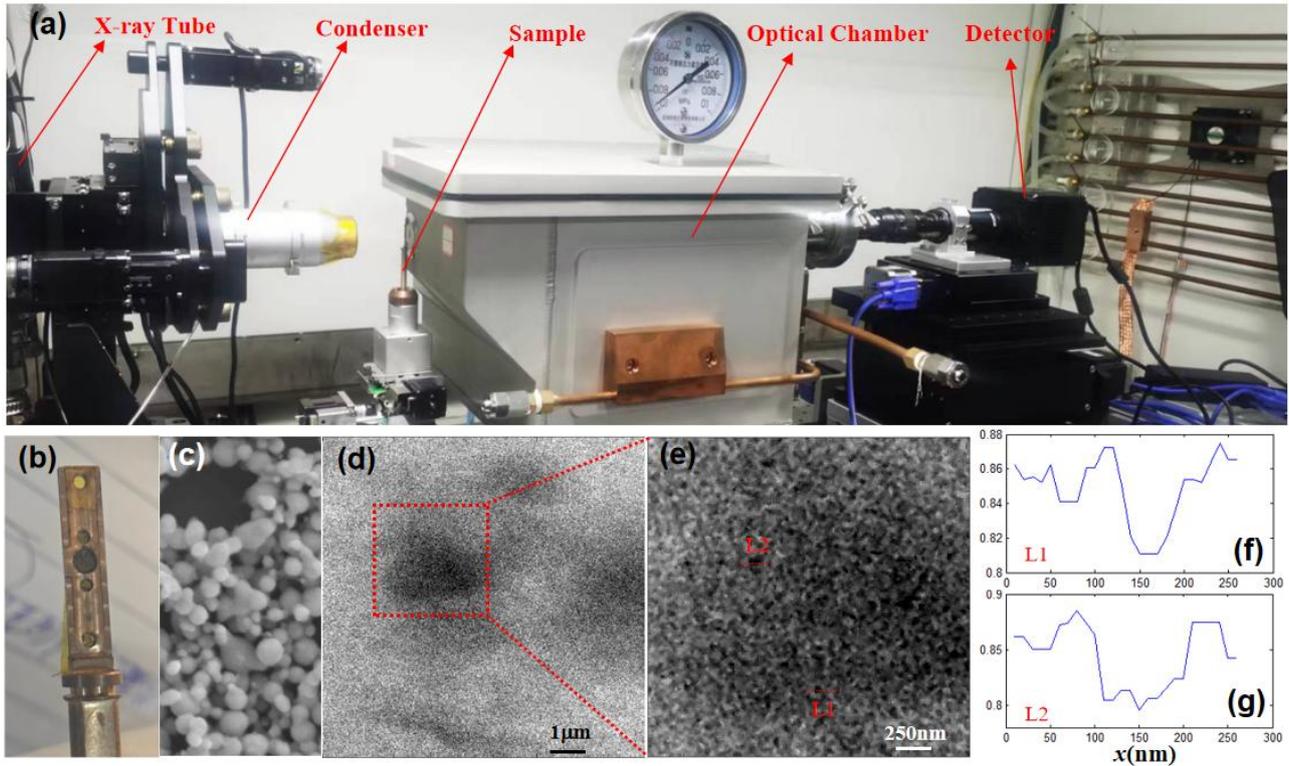

Fig 5. Preliminary imaging experiment. (a) The photograph of the TXM instrument. (b) The sample and its holder. (c) A typical SEM photograph of the sample. (d) The imaging of nanoparticles. (e) The 1D transmission curve of one selected nanoparticle.

Finally, a more accurate calculation of the spatial resolution of the TXM could be derived by using the technique of radial power spectrum density (RPSD)[9, 10]. As shown in Fig.8, the power spectrum of the sample cuts off at a spatial period of about 104nm. Therefore, the TXM spatial resolution, which is usually represented by half the spatial period could be determined as 52nm.

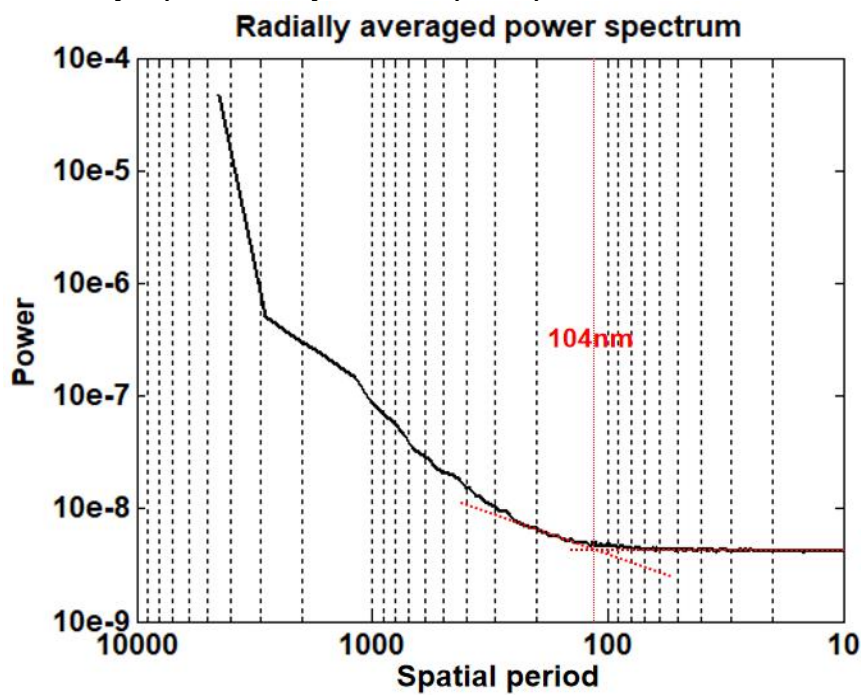

Fig 5. The radial power spectrum density (RPSD) curve.

## 4. Conclution

A high-resolution full-field transmission X-ray microscope based on the micro-focus X-ray tube was constructed. The design for this instrument was described in detail, including the optical layout, the parameters of FZP, the mechanical design and the preliminary imaging performance. The mechanical design, including X-ray source module, sample stage module, optical chamber module and detector module, was discussed briefly. Preliminary testing results gave an illumination beam about 26.1μm FWHM and a spacial resolution of 52nm. In this study, only imaging results in the absorption mode was reported. The phase contrast imaging and scattering imaging are now under development and will be presented in future work.

**Acknowledgments.**

This work is supported by the Jinan Haiyou Industry Leading Talent Project(2022), the Innovation Promotion Project of SME in Shandong Province(No.2023TSGC0093), the Enterprise Technology Innovation Project of Shandong Province(No.202350100372).